# Computer Simulation of DNA Computing-Based Boolean Matrix Multiplication

Muhammad Asad Tariq
*FAST School of Computing*
*National University of Computer & Emerging Sciences*
Lahore, Pakistan
asadtariq2002@gmail.com

Rafay Junaid
*FAST School of Computing*
*National University of Computer & Emerging Sciences*
Lahore, Pakistan
rafayjunaid@hotmail.com

Muhammad Mehdy Hasnain
*FAST School of Computing*
*National University of Computer & Emerging Sciences*
Lahore, Pakistan
mehdy922@gmail.com

Danyal Farhat
*FAST School of Computing*
*National University of Computer & Emerging Sciences*
Lahore, Pakistan
danyal.farhat@lhr.nu.edu.pk

*Abstract*—DNA computing is an unconventional approach to computing that harnesses the parallelism and information storage capabilities of DNA molecules. It has emerged as a promising field with potential applications in solving a variety of computationally complex problems. This paper explores a DNA computing algorithm for Boolean matrix multiplication proposed by Nobuyuki et al. (2006) using a computer simulation, inspired by similar work done in the past by Obront (2021) for the DNA computing algorithm developed by Adleman (1994) for solving the Hamiltonian path problem. We develop a Python program to simulate the logical operations involved in the DNA-based Boolean matrix multiplication algorithm. The simulation replicates the key steps of the algorithm, including DNA sequence generation and hybridization, without imitating the physical behaviour of the DNA molecules. It is intended to serve as a basic prototype for larger, more comprehensive DNA computing simulators that can be used as educational or research tools in the future. Through this work, we aim to contribute to the understanding of DNA-based computing paradigms and their potential advantages and trade-offs compared to conventional computing systems, paving the way for future research and advancements in this emerging field.

*Keywords—computer simulation, DNA computing, molecular computing, biological computing, Boolean matrix multiplication*

## I. INTRODUCTION

DNA computing is an unconventional approach to performing computations that harnesses the unique properties of deoxyribonucleic acid (DNA) molecules. Instead of relying on traditional electronic circuits, DNA computing uses the massive parallelism and information storage capabilities inherent in biological molecules. This emerging field was first conceived in 1994 by Leonard Adleman, who demonstrated the potential of using DNA to solve an instance of the Hamiltonian path problem.

Since Adleman's pioneering work, the field of DNA computing has witnessed significant progress, driven by the development of novel molecular biology techniques and computational models. Researchers have explored DNA-based implementations of various algorithms, including those for combinatorial optimisation problems, matrix operations, cryptography and pattern recognition tasks. However, realising practical and scalable DNA computing systems remains a formidable challenge, requiring interdisciplinary collaboration between computer scientists, biologists, and chemists.

One of the key advantages of DNA computing lies in its massive inherent parallelism. DNA molecules can undergo billions of operations simultaneously, enabled by their ability to interact and recombine based on the principles of complementary base pairing. This massive parallelism offers the potential to tackle computationally complex problems that are intractable for conventional electronic computers. Additionally, DNA's remarkable information storage density allows vast amounts of data to be encoded in a compact form, making it an attractive medium for specific computational tasks.

However, DNA computing also faces significant limitations and challenges. The process of synthesising and manipulating DNA strands can be error-prone and resource-intensive, hindering scalability. Moreover, the relatively slow rate of individual DNA operations compared to electronic circuits poses challenges for time-sensitive computations. Nonetheless, DNA computing remains an active area of research, as scientists explore innovative techniques to overcome these obstacles and leverage the unique properties of biological molecules for computational purposes. There is an ever-increasing computational demand of modern applications, due to the exponential growth of data and the complexity of problems. Conventional electronic computers have begun to approach the upper bounds of their scalability, making it crucial to explore alternative computing models that can overcome the limitations of classical systems.

In this research, we aim to simulate the working of a DNA computer for the particular task of Boolean matrix multiplication, using a Python program. Our objective is to provide an educational tool that can help newcomers to the field of DNA computing understand its fundamental principles and operations. By offering a prototype simulation, we hope to make it easier for researchers, students, and enthusiasts to grasp the basic concepts and appreciate the potential of DNA computing. This approach not only serves as a preliminary research & testing tool but also as an entry point for those interested in exploring this promising field.

Our goal is to demystify DNA computing and encourage more people to consider it as a viable area of study and research. We believe that by understanding the simulation and its underlying principles, more individuals will be inspired to contribute to the advancement of DNA computing.

The remainder of the paper is structured as follows: Section II presents an overview of related work in the field of DNA computing, highlighting key milestones and ongoing research efforts. Section III outlines the research methodology, while Section IV details the approach adopted for simulating the DNA computer. Section V presents a sample run of the simulation, as well as a brief discussion comparing DNA computers with conventional computing. Finally, Sections VI & VII conclude the paper and discuss potential future directions for this research, emphasising the importance of educational tools in promoting broader understanding and engagement in DNA computing.

## II. LITERATURE REVIEW

We begin our study of the existing literature on DNA computing with the most significant work in the area: Adleman's [1] historic experiment which laid the foundations of this field by providing a practical proof of the possibility of carrying out computations using DNA molecules. Adleman started off the paper by mentioning the possibility of building sub-microscopic computers as envisioned by Richard Feynman [2]. The problem he chose to work on to demonstrate that possibility was the well-known Hamiltonian path problem, as the exponential worst-case complexity of all its currently-known algorithms makes the problem difficult to solve even using the world's most powerful computers once the size of the graph grows large, making it a good choice for a proof-of-concept study.

In order to solve the problem and obtain a Hamiltonian path for the directed graph given in Fig. 1, Adleman employed a five-step algorithm. In the preparatory phase, he assigned a 20 base-pair long sequence of DNA to each vertex and each edge, with each half of the edges' DNA consisting of 10 bases from the linked vertices. Afterwards, he mixed together that DNA with DNA consisting of the complements of the vertices, resulting in Step 1 of the algorithm being carried out when the complement DNA molecules served to link together various sequences of edges, which lead to practically every possible path in the graph being formed. For Step 2, he used PCR (polymerase chain reaction) to tremendously amplify only the molecules representing paths that start and end at the desired source and destination vertices, effectively drowning out all other (invalid) paths. For Step 3, gel electrophoresis was carried out to filter out paths of invalid length. Step 4, on the other hand, was carried out using a system of biotin-avidin magnetic beads which ensured that only those paths which contain every vertex in the graph remained. Thus, the final product contained DNA molecules representing valid Hamiltonian paths for the given graph which were then observed in Step 5.

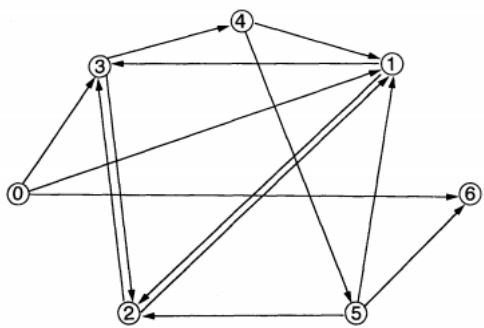

Fig. 1.
The Hamiltonian path problem considered by Adleman [1]

This experiment turned out to be monumental and opened the door to further research in the field of DNA computing, which now incorporates a variety of techniques such as DNA hybridisation, DNA origami, DNA strand displacement, dynamic DNA circuits, CRISPR, etc., many of which were not part of the original experiment. There is still plenty of room for future work on the topic, such as research into the types of algorithms that would be suitable for execution using DNA computers, as well as the potential development of a general-purpose molecular computer.

Building on Adleman's foundational work, subsequent researchers have expanded the scope of DNA computing to various computational problems, emphasising its potential beyond theoretical constructs. For instance, Nobuyuki, Ibrahim, Tsuboi, and Ono [3] presented a way to carry out Boolean matrix multiplication using DNA, building upon a prior work by Oliver [4], in which a method that uses restriction enzymes to cleave DNA molecules was proposed. Nobuyuki et al. instead proposed a method that does not require the use of restriction enzymes and is instead inspired from the method in Adleman's original paper. The idea is to represent the Boolean matrices using a directed graph. Vertices correspond to the row identifiers of the first matrix, the intermediate identifiers (columns of the first matrix and rows of the second matrix), and the column identifiers for the second matrix. This representation can be extended to form a chain of matrices as well. Edges exist between vertices representing non-zero elements.

For the DNA representation of the vertices and edges, they used practically the same mechanism as [1]. After preparing the DNA molecules, they carried out the experiment by using POA (Parallel Overlap Assembly) to generate DNA strands representing all possible paths in the graph. Then the resulting solution was distributed amongst as many test tubes as there are elements in the product matrix. Within each test tube, they added primers to identify the existence of required paths (those that connect initial, intermediate and terminal vertices with each other, because they represent non-zero values in the product matrix) using PCR. Once the test tubes containing the paths were identified, polyacrylamide gel electrophoresis was used to visualise the results of the computation, with the bands on the gel representing the output of the process.

The experimenters multiplied a chain of four 2 by 2 matrices but indicated their intention of continuing the work in the future and extending it to the multiplication of 10 by 10 matrices using molecular computers. Perhaps another direction of future work could be incorporating non-Boolean matrices to the technique. Moreover, comparative analysis of the relative performance of DNA computing-based versus classical matrix multiplication algorithms would be essential in order to determine how beneficial DNA computing actually can be in solving practical problems.

Moving on, Watada and Binti Abu Bakar [5] conducted extensive research on the various applications that DNA or molecular computing has to offer. The initial portion of their paper reiterated Adleman's work [1] as discussed above, followed by some discussion on the quantitative advantages that DNA computing offers in comparison to conventional supercomputers, i.e. it is expected to outshine our current systems by several orders of magnitude in both the number of parallel operations performed per second as well as offering lower energy consumption rates.

Furthermore, they divide the research done up till 2008 into two categories: biochemical feasibility and engineering/application problems. They elaborated how DNA computing research has seen advancements in both encoding numerical values and tackling real-world problems. Scientists have explored various methods, including using DNA strand length or melting temperature to represent numbers. These techniques have been applied to solve problems like scheduling, cryptography, and forecasting. However, efficiently representing numbers and achieving broader adoption remain key challenges in the field. Another problem lies in how gel electrophoresis, the standard method for analysing DNA strands, suffers from limitations. Additionally, traditional PCR, a technique used to amplify DNA for analysis, has limitations that can affect results. Researchers are addressing these limitations by exploring real-time PCR, which allows for real-time monitoring of the experiment, improving the readability of end results.

Even from the small part of the literature reviewed above, it is obvious that the field of DNA computing has much to offer. However, it appears to be a rather obscure area to many. Effort needs to be made to draw more students and researchers to this field in order to fully realise its considerable potential. An effort in this regard was made by Obront [6] using a basic Python program to simulate the original DNA computing experiment detailed in [1] in order to make it easier to understand for people who are new to the field. This was the inspiration for our own work that follows.

III. METHODOLOGY

Now, we state the methodology we used for our work in simulating a DNA computing algorithm. As stated earlier, our objective was to simulate the working of a DNA computing algorithm, similar to the way it was done in [6] for the algorithm from [1]. For this purpose, we reviewed the literature on DNA-based algorithms, and the one we chose for our simulation was the Boolean matrix multiplication algorithm presented in [3]. It is important to note, though, that the simulation is supposed to be a *logical* one, and thus does not simulate the *physical* or *chemical* interactions of the DNA molecules due to the limit of its scope. As a matter of fact, constructing a more advanced version of the simulator having the capability to simulate the chemical reactions that are carried out during the course of computation using DNA could be a natural extension of this work in the future.

Let us first discuss the main objectives and uses of this work, followed by an overview of the process.

A. *Objectives:*

1) *As an educational tool:*

One of the main aims of our work is to bring DNA computing algorithms into the notice of more people from the field of computer science. In this regard, the simulation can allow people not related to the field of biological computing but familiar with programming basics to understand DNA algorithms, making the field much more widely accessible. Rather than focusing solely on performance comparisons with classical computing systems, our approach emphasizes the educational value of the simulation. We believe that by providing a hands-on experience with DNA algorithms, individuals from diverse backgrounds can gain a better appreciation for the unique strengths and potential applications of DNA computing.

2) *DNA algorithm development and testing:*

The program can be used as a blueprint for developing new algorithms in code and test their functionality before actually carrying out the experiment practically, which might be difficult due to resource constraints. This use case might not be applicable to the simulation as it currently stands, but if some work is carried out to enhance the capabilities of the program, then it can certainly serve as a useful tool for researchers as well, in addition to its use in education.

B. *Process:*

We begin with an overview of the Boolean matrix multiplication algorithm using an example given in [3]. The two matrices to be multiplied are:

$$X = \begin{pmatrix} 0 & 1 \\ 1 & 0 \end{pmatrix} \text{ and } Y = \begin{pmatrix} 0 & 1 \\ 1 & 0 \end{pmatrix}.$$

According to the usual rules for matrix multiplication, the product of these matrices is:

$$Z = XY = \begin{pmatrix} 0 \cdot 0 + 1 \cdot 1 & 0 \cdot 1 + 1 \cdot 0 \\ 1 \cdot 0 + 0 \cdot 1 & 1 \cdot 1 + 0 \cdot 0 \end{pmatrix} = \begin{pmatrix} 1 & 0 \\ 0 & 1 \end{pmatrix}$$

In order to carry out this multiplication using the DNA-based algorithm, we first create two vertices corresponding to the rows of the first matrix X (let us call them 1 and 2), two vertices corresponding to the columns of the first matrix which also correspond to the rows of the second matrix Y (let us call them '$a$' and '$b$'), and two matrices corresponding to the columns of the second matrix (let us call them '$A$' and '$B$'). Next, we set up edges between those vertices which represent the position of a 1-value in either matrix. For example, we have a value of 1 in the first row and second column of matrix X, so we set up a directed edge between vertices 1 and 'b', as shown in Fig. 2 below.

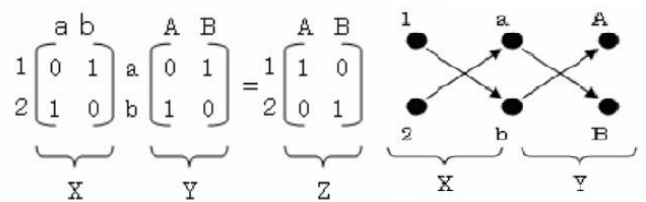

Fig. 2.
A product of two Boolean matrices and its graphical representation [3]

As shown by the authors of [3], this reduces the problem of multiplying the Boolean matrices to finding the possible paths between the vertices representing the rows of the first matrix and the ones representing the columns of the last matrix (which, in this case, is the second matrix, but the algorithm is generalisable to a chain consisting of an arbitrary number of Boolean matrices being multiplied with each other, as long as they are compatible for matrix multiplication, i.e. the number of columns in the matrix on the left is equal to the number of rows in the matrix on the right), which indicate the 1-values in the product matrix.

IV. IMPLEMENTATION

The example we will be using to demonstrate the working of our simulation is, in fact, the one that was actually solved via DNA computing in [3]. Let us now move to the code (the latest version of which will be maintained at https://github.com/ASD0x41/dna-computing-simulator).

```
1)   import random
2)   base_pairs = {'A': 'T', 'T': 'A', 'C': 'G', 'G': 'C'}
```

The *random* module will be used for generating random DNA sequences to represent the vertices of the graph while *base_pairs* maps DNA bases to their complements (i.e. adenine (A) to thymine (T), cytosine (C) to guanine (G) and vice versa).

```
3)   strand_length = 10
4)
5)   matrices = [
6)       [[0, 1],  # a
7)        [1, 0]   # b
8)       # c  d],
9)       [[1, 0],  # c
10)       [0, 1]   # d
11)      # e  f],
12)      [[0, 1],  # e
13)       [1, 0]   # f
14)      # g  h],
15)      [[1, 0],  # g
16)       [0, 1]   # h
17)      # i  j]
18)  ]
19)
20)  vertex_labels = {
21)      (0, 0): 'a', (0, 1): 'b',
22)      (1, 0): 'c', (1, 1): 'd',
23)      (2, 0): 'e', (2, 1): 'f',
24)      (3, 0): 'g', (3, 1): 'h',
25)      (4, 0): 'i', (4, 1): 'j'
26)  }
```

The above segment represents the inputs to the simulation. We have four 2 by 2 Boolean matrices, and we have assigned labels to each vertex for convenience and compatibility with the notation used in [3].

```
27)  assert strand_length % 2 == 0
28)
29)  matrix_count = len(matrices)
30)  assert matrix_count > 1
31)
32)  assert all(len(matrices[i]) > 0 for i in
     range(matrix_count))
33)  for i in range(matrix_count):
34)      assert all(len(matrices[i][j]) > 0 for j in
     range(len(matrices[i])))
35)      for j in range(len(matrices[i])):
36)          assert all(not isinstance(matrices[i][j][k],
     list) for k in range(len(matrices[i][j])))
37)          assert all((matrices[i][j][k] in (0, 1)) for
     k in range(len(matrices[i][j])))
38)  assert all(len(set(len(matrices[i][j]) for j in
     range(len(matrices[i])))) == 1 for i in
     range(matrix_count))
39)  assert all(len(matrices[i - 1][0]) ==
     len(matrices[i]) for i in range(1, matrix_count))
40)
41)  vertex_count = len(matrices[0]) +
     sum(len(matrices[i]) for i in range(1, matrix_count))
     + len(matrices[matrix_count - 1][0])
42)  initial_vertices, terminal_vertices =
     len(matrices[0][0]), len(matrices[-1][0])
43)  assert len(vertex_labels) == vertex_count
44)  assert all(isinstance(key, tuple) and len(key) == 2
     and isinstance(key[0], int) and isinstance(key[1],
     int) for key in vertex_labels.keys())
45)  for i in range(matrix_count + 1):
46)      assert all((i, j) in vertex_labels.keys() for j
     in range(len(matrices[i]) if i < matrix_count else
     terminal_vertices))
47)  assert all(isinstance(value, str) and len(value) > 0
     for value in vertex_labels.values())
```

The above code segment is not directly related to the simulation logic. Instead, it is there for input validation, so that if someone attempts to run the code while providing invalid inputs, an assertion failure can identify the issue. The checks are related to the structure and content of the *matrices* list and the *vertex_labels* dictionary.

In addition, the parity of *strand_length* needs to be even so that edge strands can be formed from halves of the linked vertex strands.

```
48)  def GenerateRandomDNASequence(sequence_length):
49)      return
     ''.join(random.choices(list(base_pairs.keys()), k =
     sequence_length))
50)
51)  vertex_strands = []
52)  vertex_encodings = {}
53)  vertex_decodings = {}
54)
55)  for i in range(matrix_count + 1):
56)      vertex_layer = []
57)      for j in range(len(matrices[i]) if i <
     matrix_count else terminal_vertices):
58)          new_strand =
     GenerateRandomDNASequence(strand_length)
59)
60)          vertex_layer.append(new_strand)
61)          vertex_encodings[vertex_labels[(i, j)]] =
     new_strand
62)          vertex_decodings[new_strand] = (i, j)
63)      vertex_strands.append(vertex_layer)
64)
65)  print("Vertex Strands:")
66)  for vertex_label, vertex_strand in
     vertex_encodings.items():
67)      print("\t" + vertex_label + " : " +
     vertex_strand)
```

In the above part, we generate random DNA sequences to represent the vertices of the graph. This is followed by generating the appropriate sequences for the edges (consisting of complements of halves of the strands representing vertices connected by a particular edge) below:

```
68)  def GenerateComplementaryDNASequence(sequence):
69)      return ''.join(base_pairs[sequence[i]] for i in
     range(len(sequence)))
70)  edge_strands = []
71)  edge_encodings = {}
72)  edge_decodings = {}
73)
74)  for i in range(matrix_count):
75)      for j in range(len(matrices[i])):
76)          for k in range(len(matrices[i][j])):
77)              if matrices[i][j][k] == 1:
78)                  left_half =
     GenerateComplementaryDNASequence(vertex_strands[i][j]
     [strand_length // 2:])
79)                  right_half =
     GenerateComplementaryDNASequence(vertex_strands[i +
     1][k][:strand_length // 2])
80)                  new_strand = left_half + right_half
81)
82)                  edge_strands.append(new_strand)
83)                  edge_encodings[(vertex_labels[(i,
     j)], vertex_labels[(i + 1, k)])] = new_strand
84)                  edge_decodings[new_strand] =
     (vertex_labels[(i, j)], vertex_labels[(i + 1, k)])
85)
86)  print("Edge Strands:")
87)  for edge_strand, edge_label in
     edge_decodings.items():
88)      print("\t(" + edge_label[0] + "," + edge_label[1]
     + ") : " + edge_strand)
```

Next, we start with the set of vertices, and begin growing the strands. If there is an edge connecting the end of the strand with another vertex, we append that vertex to the end of the strand, which represents a possible path in the graph. In this case, we are maintaining only one copy of each path. However, in reality, there are innumerable copies of DNA strands representing each possible path in the template solution. The sheer number of those copies ensures that practically all possibles paths will be enumerated despite the randomness in the interactions of the DNA molecules. Nevertheless, we treat this part differently in the code.

```
89)     template = set()
90)     for i in range(matrix_count + 1):
91)         for j in range(len(matrices[i]) if i <
    matrix_count else terminal_vertices):
92)             template.add(vertex_strands[i][j])
93)
94)     def Match(strandOne, strandTwo):
95)         strand_len = len(strandOne)
96)         return strand_len == len(strandTwo) and
    all(strandOne[i] == base_pairs[strandTwo[i]] for i in
    range(strand_len))
97)
98)     for _ in range(matrix_count):
99)         growing_strands = set()
100)        for path_strand in template:
101)            for edge_strand in edge_strands:
102)                if Match(path_strand[-(strand_length //
    2):], edge_strand[:strand_length // 2]):
103)                    growing_strands.add(path_strand + "-"
    + vertex_encodings[edge_decodings[edge_strand][1]])
104)        template = template.union(growing_strands)
105)
106)    print("Path Strands:")
107)    for path_strand in template:
108)        print("\t" + path_strand + " (" +
    vertex_labels[vertex_decodings[path_strand[:strand_le
    ngth]]], end = '')
109)        for i in range(1, (len(path_strand) + 1) //
    (strand_length + 1)):
110)            print(" -> " +
    vertex_labels[vertex_decodings[path_strand[i *
    (strand_length + 1) : (i + 1) * (strand_length + 1) -
    1]]], end = '')
111)        print(")")
```

We now have a set containing DNA sequences for all possible paths in the graph. What is left is to identify the paths that start from an initial vertex and end at a terminal vertex, representing the elements of the product matrix.

```
112)    print("Complete Paths:")
113)
114)    product_matrix = [[0 for j in
    range(terminal_vertices)] for i in
    range(initial_vertices)]
115)
116)    for path_strand in template:
117)        starting_vertex = path_strand[:strand_length]
118)        ending_vertex = path_strand[-strand_length:]
119)
120)        if starting_vertex in vertex_strands[0] and
    ending_vertex in vertex_strands[matrix_count]:
121)            product_matrix[vertex_decodings[starting_vert
    ex][1]][vertex_decodings[ending_vertex][1]] = 1
122)
123)            print("\t" + path_strand + " (" +
    vertex_labels[vertex_decodings[path_strand[:strand_le
    ngth]]], end = '')
124)            for i in range(1, (len(path_strand) + 1) //
    (strand_length + 1)):
125)                print(" -> " +
    vertex_labels[vertex_decodings[path_strand[i *
    (strand_length + 1) : (i + 1) * (strand_length + 1) -
    1]]], end = '')
126)            print(")")
127)
128)    print("\nProduct Matrix:")
129)    print("\t", product_matrix)
```

Finally, we decode the product matrix from the filtered paths. Those elements for which a strand starting from the appropriate and ending at the appropriate initial terminal vertices exists have a value of 1 while all the rest are zeros.

## V. RESULTS

For the inputs provided in the previous section, the program gives the following output on the console for one sample run (the DNA sequences produced will be different each time the program is executed due to the fact that they are generated randomly; however, the results are supposed to be the same for a particular set of inputs):

```
Vertex Strands:
    a : ATTCGTTCTT
    b : TCTTGGACAG
    c : GCACAGTTAA
    d : ATCACACAAT
    e : GCCCTTCTGG
    f : TAAAACGGCT
    g : GCTTGGACAG
    h : CTCAGTGAGC
    i : GTCATGAGCG
    j : ATGAACGGTG

Edge Strands:
    (a,d) : AAGAATAGTG
    (b,c) : CTGTCCGTGT
    (c,e) : CAATTCGGGA
    (d,f) : TGTTAATTTT
    (e,h) : AGACCGAGTC
    (f,g) : GCCGACGAAC
    (g,i) : CTGTCCAGTA
    (h,j) : ACTCGTACTT

Path Strands:
    GCTTGGACAG-GTCATGAGCG (g -> i)
    GCCCTTCTGG (e)
    TCTTGGACAG-GCACAGTTAA-GCCCTTCTGG (b -> c -> e)
    TAAAACGGCT (f)
    TAAAACGGCT-GCTTGGACAG-GCACAGTTAA-GCCCTTCTGG (f -> g -> c -> e)
    GCCCTTCTGG-CTCAGTGAGC (e -> h)
    CTCAGTGAGC-ATGAACGGTG (h -> j)
    GCACAGTTAA-GCCCTTCTGG-CTCAGTGAGC-ATGAACGGTG (c -> e -> h -> j)
    ATTCGTTCTT-ATCACACAAT-TAAAACGGCT-GCTTGGACAG-GTCATGAGCG (a -> d -> f -> g -> i)
    GCTTGGACAG (g)
    TCTTGGACAG-GCACAGTTAA (b -> c)
    ATTCGTTCTT-ATCACACAAT-TAAAACGGCT (a -> d -> f)
    GCTTGGACAG-GCACAGTTAA (g -> c)
    GCACAGTTAA (c)
    ATTCGTTCTT-ATCACACAAT-TAAAACGGCT-GCTTGGACAG-GCACAGTTAA (a -> d -> f -> g -> c)
    TCTTGGACAG-GCACAGTTAA-GCCCTTCTGG-CTCAGTGAGC-ATGAACGGTG (b -> c -> e -> h -> j)
    TCTTGGACAG (b)
    ATTCGTTCTT (a)
    ATCACACAAT-TAAAACGGCT-GCTTGGACAG-GCACAGTTAA-GCCCTTCTGG (d -> f -> g -> c -> e)
    TAAAACGGCT-GCTTGGACAG-GCACAGTTAA-GCCCTTCTGG-CTCAGTGAGC (f -> g -> c -> e -> h)
    TCTTGGACAG-GTCATGAGCG (b -> i)
    GCTTGGACAG-GCACAGTTAA-GCCCTTCTGG (g -> c -> e)
    TAAAACGGCT-GCTTGGACAG (f -> g)
    GCACAGTTAA-GCCCTTCTGG-CTCAGTGAGC (c -> e -> h)
    CTCAGTGAGC (h)
    ATTCGTTCTT-ATCACACAAT (a -> d)
    GCTTGGACAG-GCACAGTTAA-GCCCTTCTGG-CTCAGTGAGC (g -> c -> e -> h)
    ATCACACAAT-TAAAACGGCT (d -> f)
    ATCACACAAT-TAAAACGGCT-GCTTGGACAG-GCACAGTTAA (d -> f -> g -> c)
    TAAAACGGCT-GCTTGGACAG-GTCATGAGCG (f -> g -> i)
    GCCCTTCTGG-CTCAGTGAGC-ATGAACGGTG (e -> h -> j)
    ATTCGTTCTT-ATCACACAAT-TAAAACGGCT-GCTTGGACAG (a -> d -> f -> g)
    TAAAACGGCT-GCTTGGACAG-GCACAGTTAA (f -> g -> c)
    ATCACACAAT-TAAAACGGCT-GCTTGGACAG (d -> f -> g)
    TCTTGGACAG-GCACAGTTAA-GCCCTTCTGG-CTCAGTGAGC (b -> c -> e -> h)
    GCTTGGACAG-GCACAGTTAA-GCCCTTCTGG-CTCAGTGAGC-ATGAACGGTG (g -> c -> e -> h -> j)
    GTCATGAGCG (i)
    ATCACACAAT (d)
    GCACAGTTAA-GCCCTTCTGG (c -> e)
    ATGAACGGTG (j)
    ATCACACAAT-TAAAACGGCT-GCTTGGACAG-GTCATGAGCG (d -> f -> g -> i)

Complete Paths:
    ATTCGTTCTT-ATCACACAAT-TAAAACGGCT-GCTTGGACAG-GTCATGAGCG (a -> d -> f -> g -> i)
    TCTTGGACAG-GCACAGTTAA-GCCCTTCTGG-CTCAGTGAGC-ATGAACGGTG (b -> c -> e -> h -> j)
    TCTTGGACAG-GTCATGAGCG (b -> i)

Product Matrix:
    [[1, 0], [1, 1]]
```

It can be seen that the algorithm worked by finding paths between the vertices representing the rows of the first matrix (i.e. '$a$' and '$b$') and the ones representing the columns of the last matrix (i.e. '$i$' and '$j$'), which are $a \rightarrow d \rightarrow f \rightarrow g \rightarrow i$ and $b \rightarrow c \rightarrow e \rightarrow h \rightarrow j$. As a result, the elements that have a value of 1 in the product matrix are those that are in the first row, first column (due to '$a$' and '$i$') and in the second row, second column (due to '$b$' and '$j$'). This is in line with the results obtained by practically carrying out the DNA computation as detailed in [3] and shown in Fig. 3. As an aside, however, one difference between the above implementation and the one in [3] is that we have used a strand length of 10 instead of 20 in order to keep the length of the output limited.

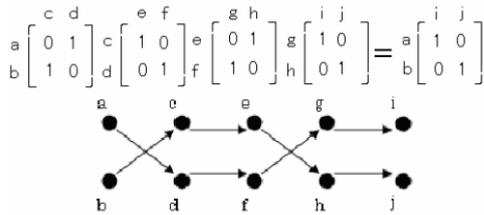

Fig. 3.
The product of four 2 by 2 Boolean matrices computed in [3]

As stated earlier, the purpose of this simulation is to act as an educational tool for enhancing understanding of DNA computing. In this regard, we present a few points regarding the vast potential of this novel computing approach.

In Adleman's seminal work [1], he proposed scaling his setup to perform 10^20 operations per second, surpassing even the fastest supercomputer at the time, which managed 10^12 operations per second. Currently in 2024, the fastest supercomputer offers 10^18 operations per second, demonstrating the high potential of DNA computing. Moreover, Adleman demonstrated the exponential scalability of DNA computing. For instance, multiplying matrices in sizes of billions would still require just a single test tube of DNA. In contrast, conventional digital computers face significant scalability limitations due to limited hardware resources.

DNA computing offers energy efficiency driven by chemical reactions, requiring relatively little energy compared to traditional electronic computers. While detailed energy studies are lacking, the inherent low-power nature of biological processes suggests substantial energy savings, particularly for large-scale computations. Furthermore, as DNA manufacturing technology advances, the overhead involved is expected to be reduced further. Additionally, DNA computing minimizes data transfer overhead since all data is simultaneously available for parallel computations. This makes DNA computing a promising avenue for research across various computing domains.

## VI. Conclusion

In this study, we developed a computer simulation of a DNA computing-based Boolean matrix multiplication algorithm, as proposed by Nobuyuki et al. Through our Python-based simulation, we replicated the key steps of the algorithm to demonstrate the use of DNA computing for matrix operations.

While our simulation does not capture the physical behaviour of DNA molecules, it serves as an educational tool to help researchers and students understand the fundamental principles of DNA computing.

Our work underscores the potential advantages of DNA computing, such as massive parallelism and high information storage density, which can make it a valuable alternative to conventional electronic computing for solving computationally complex problems. However, we also acknowledge the significant challenges associated with DNA computing, including the error-prone nature of DNA manipulation and the slow rate of DNA operations, though those are considerably offset by the gain from parallelism.

The results of our simulation suggest that DNA computing holds promise for specific applications, particularly those that can leverage its parallel processing capabilities. Our goal is to contribute to the broader understanding and appreciation of DNA computing, encouraging more researchers, students, and enthusiasts to explore its potential. By providing a prototype simulation, we hope to inspire future advancements and innovations in DNA computing, paving the way for the development of practical and scalable molecular computing systems.

## VII. Future Directions

With regard to the development of a DNA computing simulator, there is much that can be done by building upon this work. Currently, our simulation is closely associated with one particular algorithm, which uses only a small subset of all the techniques that are used in the field. By generalising the scope of the program, a more comprehensive simulation having the capabilities to simulate a much larger variety of DNA-based algorithms involving more advanced DNA generation, manipulation and analysis techniques could be developed. Furthermore, work could be done on providing a better interface, perhaps a GUI, in order to help users better visualise what goes on during the course of a DNA computation and interact with the simulator more conveniently. Similarly, the simulator could be enhanced by incorporating the physical aspects of DNA computations as well as its logical ones. Thus, there is ample scope for future work in this aspect, which could help make such simulations more useful to educators and researchers.

## References

[1] L. M. Adleman, "Molecular Computation of Solutions to Combinatorial Problems," *New Series Science*, vol. 266, issue 5187, pp. 1021-1024, November 1994.

[2] R. P. Feynman, "Miniturization," *D. H. Gilbert, Ed.* pp. 282-296, 1961.

[3] K. Nobuyuki, Z. Ibrahim, Y. Tsuboi & O. Ono, "Matrix Multiplication with DNA Computing," in *The Eleventh International Symposium on Artificial Life and Robotics*, Oita, Japan, January, 2006.

[4] J. S. Oliver, "Matrix multiplication with DNA," *Journal of Molecular Evolutioni*, vol. 45, issue 2, pp. 161-167, August 1997.

[5] J. Watada, R Binte Abu Bakar, "DNA Computing and Its Applications," in *Eighth International Conference on Intelligent Systems Design and Applications*, Kitakyushu, Japan, 2008.

[6] Z. Obront, "dna-computing-simulator," July, 2021. [Online]. Available: https://github.com/zobront/dna-computing-simulator. [Accessed May 30, 2024].